\documentclass[12pt]{article}
\usepackage{amsfonts}
\usepackage{epsf}

\usepackage{latexsym}

\newcommand{\beq}{\begin{equation}}
\newcommand{\eeq}{\end{equation}}
\newcommand{\beqs}{\begin{eqnarray}}
\newcommand{\eeqs}{\end{eqnarray}}
\newcommand{\reff}[1]{(\ref{#1})}

\catcode`@=11
\@addtoreset{equation}{section}
\@addtoreset{equation}{subsection}
\def\theequation{\ifnum\value{section}=0 \arabic{equation}\ignorespaces
\else \ifnum\value{section}=-1 A.\arabic{equation}\ignorespaces
\else \ifnum\value{subsection}=0 \thesection.\arabic{equation}\ignorespaces
\else \thesection.\arabic{subsection}.\arabic{equation}\ignorespaces
                           \fi
                      \fi
                 \fi}
\catcode`@=12

\begin{document}

\def\thefootnote{\fnsymbol{footnote}}

\baselineskip 6.0mm

\vspace{4mm}

\begin{center}

{\Large \bf Exact $T=0$ Partition Functions for Potts Antiferromagnets on
Sections of the Simple Cubic Lattice} 

\vspace{8mm}

\setcounter{footnote}{0}
Jes\'us Salas$^{(a)}$\footnote{email: jesus@melkweg.unizar.es} and
\setcounter{footnote}{6}
Robert Shrock$^{(b)}$\footnote{email: robert.shrock@sunysb.edu}

\vspace{6mm}

(a) \ Departamento de F\'{\i}sica Te\'orica \\
Facultad de Ciencias \\ 
Universidad de Zaragoza \\
Zaragoza 50009 \\
Spain

(b) \ C. N. Yang Institute for Theoretical Physics  \\
State University of New York       \\
Stony Brook, N. Y. 11794-3840  \\
USA

\vspace{10mm}

{\bf Abstract}
\end{center}

We present exact solutions for the zero-temperature partition function of the
$q$-state Potts antiferromagnet (equivalently, the chromatic polynomial $P$) on
tube sections of the simple cubic lattice of fixed transverse size $L_x \times
L_y$ and arbitrarily great length $L_z$, for sizes $L_x \times L_y = 2 \times
3$ and $2 \times 4$ and boundary conditions (a) $(FBC_x,FBC_y,FBC_z)$ and (b)
$(PBC_x,FBC_y,FBC_z)$, where $FBC$ ($PBC$) denote free (periodic) boundary
conditions. In the limit of infinite-length, $L_z \to \infty$, we calculate the
resultant ground state degeneracy per site $W$ (= exponent of the ground-state
entropy).  Generalizing $q$ from ${\mathbb Z}_+$ to ${\mathbb C}$, we determine
the analytic structure of $W$ and the related singular locus ${\cal B}$ which
is the continuous accumulation set of zeros of the chromatic polynomial.  For
the $L_z \to \infty$ limit of a given family of lattice sections, $W$ is
analytic for real $q$ down to a value $q_c$.  We determine the values of $q_c$
for the lattice sections considered and address the question of the value of
$q_c$ for a $d$-dimensional Cartesian lattice.  Analogous results are presented
for a tube of arbitrarily great length whose transverse cross section is formed
from the complete bipartite graph $K_{m,m}$.

\vspace{16mm}

\pagestyle{empty}
\newpage

\pagestyle{plain}
\pagenumbering{arabic}
\renewcommand{\thefootnote}{\arabic{footnote}}
\setcounter{footnote}{0}

\section{Introduction} 
\label{sec_intro}

The $q$-state Potts antiferromagnet (PAF) \cite{potts,wurev} exhibits nonzero
ground state entropy, $S_0 > 0$ (without frustration) for sufficiently large
$q$ on a given lattice $\Lambda$ or, more generally, on a graph $G$.  This is
equivalent to a ground state degeneracy per site $W > 1$, since $S_0 = k_B \ln
W$.  Such nonzero ground state entropy is important as an exception to the
third law of thermodynamics \cite{al,cw}.  One physical example is provided by
ice, for which the residual molar entropy is $S_0 = 0.82 \pm 0.05$
cal/(K-mole), i.e., $S_0/R = 0.41 \pm 0.03$, where $R=N_{Avog.}k_B$ \cite{lp}.
Indeed, residual entropy at low temperatures has been observed in a number of
molecular crystals, including nitrous oxide, NO and FClO$_3$ (a comprehensive
review is given in Ref. \cite{ps}).  In these physical examples, the entropy
occurs without frustration, i.e., the configurational energy can be minimized,
just as in the Potts antiferromagnet for sufficiently large $q$.

There is a close connection with graph theory here, since the zero-temperature
partition function of the above-mentioned $q$-state Potts antiferromagnet on a
graph $G=(V,E)$ satisfies
\beq
Z(G,q,T=0)_{\rm PAF}=P(G,q)
\label{zp}
\eeq
where $G$ is defined by its set of vertices $V$ and edges $E$ and
$P(G,q)$ is the chromatic polynomial expressing the number of ways of
coloring the vertices of $G$ with $q$ colors such that no two
adjacent vertices have the same color (for reviews, see
\cite{rrev}-\cite{bbook}).  The minimum number of colors necessary for such a
coloring of $G$ is called the chromatic number, $\chi(G)$. 
Thus
\beq
W(\{G\},q) = \lim_{n \to \infty} P(G,q)^{1/n}
\label{w}
\eeq 
where $n=|V|$ is the number of vertices of $G$ and we denote the formal
infinite-length limit of strip graphs of type $G$ as 
$\{G\} = \lim_{n \to \infty}G$.  At certain special points $q_s$ 
(typically $q_s=0,1,.., \chi(G)$), one has the noncommutativity of
limits \cite{w} 
\beq
\lim_{q \to q_s} \lim_{n \to \infty} P(G,q)^{1/n} \ne \lim_{n \to
\infty} \lim_{q \to q_s}P(G,q)^{1/n}
\label{wnoncom}
\eeq 
and hence it is necessary to specify the order of the limits in the definition
of $W(\{G\},q_s)$. Denoting $W_{qn}$ and $W_{nq}$ as the functions defined by
the different order of limits on the left and right-hand sides of
(\ref{wnoncom}), we take $W \equiv W_{qn}$ here; this has the advantage of
removing certain isolated discontinuities that are present in $W_{nq}$. Using
the expression for $P(G,q)$, one can generalize $q$ from ${\mathbb Z}_+$ to
${\mathbb C}$.  The zeros of $P(G,q)$ in the complex $q$ plane are called
chromatic zeros; a subset of these may form an accumulation set in the $n \to
\infty$ limit, denoted ${\cal B}$, which is the continuous locus of points
where $W(\{G\},q)$ is nonanalytic. For some families of graphs ${\cal B}$ may
be null, and $W$ may also be nonanalytic at certain discrete points.  The
maximal region in the complex $q$ plane to which one can analytically continue
the function $W(\{G\},q)$ from physical values where there is nonzero ground
state entropy is denoted $R_1$.  The ground state degeneracy per site
$W(\{G\})$ is an analytic function of real $q$ from large values down to the
value $q_c$, which is the maximal value where ${\cal B}$ intersects the
(positive) real axis. For some families of graphs, ${\cal B}$ does not cross or
intersect the real $q$ axis; in these cases, no $q_c$ is defined. However, even
in cases where no such intersection occurs, ${\cal B}$ often includes
complex-conjugate arcs with endpoints close to the positive real axis, and
hence, in these cases, it can be useful to define a quantity $(q_c)_{\rm eff}$
equal to the real part of the endpoints.  We shall use this definition here.

In this work we present exact solutions for chromatic polynomials
$P(G,q)$ for sections of the simple cubic (sc) lattice with fixed transverse
size $L_x \times L_y$ and arbitrarily great length $L_z$, for cross sections
$L_x \times L_y = 3 \times 2$ and $4 \times 2$. These calculations are carried
out for the cases (a) $(FBC_x,FBC_y,FBC_z)$ (rectangular solid) and (b)
$(PBC_x,FBC_y,FBC_z)$ (homeomorphic to an annular cylindrical solid), where
$FBC_i$ and $PBC_i$ denote free and periodic boundary conditions in the $i$'th
direction, respectively.  We shall use the notation $(L_i)_{\rm F}$ and
$(L_i)_{\rm P}$ to denote free and periodic boundary conditions in the $i$'th
direction, so that, for example, the $3 \times 2 \times L_z$ sections of the
simple cubic lattice with the boundary conditions of type (a) and (b) are
denoted $3_{\rm F} \times 2_{\rm F} \times (L_z)_{\rm F}$ and $3_{\rm P} \times
2_{\rm F} \times (L_z)_{\rm F}$, respectively.  For each family of graphs,
taking the infinite-length limit $L_z \to \infty$, we calculate $W(\{G\},q)$,
${\cal B}$, and hence $q_c$.

We also present corresponding results for a tube of arbitrarily great 
length whose transverse cross section is formed from the complete bipartite 
graph $K_{m,m}$, for the cases $m=2$ and 3.  Here the complete graph $K_n$ is
defined as the graph consisting of $n$ vertices such that each vertex is
connected by edges (bonds) to every other vertex, and the complete bipartite
graph $K_{m,n}$ is defined as the join $K_m + K_n$, where the join of two
graphs $G$ and $H$, denoted $G+H$ is the graph obtained by joining each of the
vertices of $G$ to each of the vertices of $H$.  Although the complete
bipartite graphs are not regular lattices as studied in statistical mechanics, 
they are useful since they allow us to obtain exact results for cases of high 
effective coordination number.

There are several motivations for this work.  We have mentioned the basic
importance of nonzero ground state entropy in statistical mechanics and
physical examples of this phenomenon. From the point of view of rigorous
statistical mechanics, exact analytic solutions are always valuable since they
complement results from approximate series expansions and numerical methods.
We have defined the point $q_c$ above in terms of the function $W(\{G\},q)$.
This point has another important physical significance: for the $n \to \infty$
limit of a given family of graphs, $\{G\}$, the $q$-state Potts antiferromagnet
has no finite-temperature phase transition but is disordered for all $T \ge 0$
if $q > q_c(\{G\})$, and has a zero-temperature critical point for
$q=q_c(\{G\})$ \cite{qcni}. 
 For the Potts model on the (infinite) square lattice,
via a mapping to a vertex model, it has been concluded that $q=3$
\cite{lenard}.  However, the value of $q_c$ is not known for any lattice of
dimension three or higher.  One of the main motivations for our study is the
insight that one gains concerning the dependence of $q_c$ on the coordination
number $\Delta=2d$ of a $d$-dimensional Cartesian lattice ${\mathbb E}^d$ for
the case $d=3$. Furthermore, although infinite-length sections of
higher-dimensional lattices with fixed finite $(d-1)$-dimensional volume
transverse to the direction in which the length goes to infinity are
quasi-one-dimensional systems and hence (for finite-range spin-spin
interactions) do not have finite-temperature phase transitions, their
zero-temperature critical points are of interest.  Finally, in addition to the
physics motivations, the present results are of interest in mathematical graph
theory.  Some related earlier work is in 
Refs.~\cite{ssbounds}-\cite{JS}.

Our exact calculations of $q_c$ for the infinite-length limits of tube sections
of the simple cubic lattice yield information relevant to estimates of $q_c$
for the infinite cubic lattice.  The point here is that as the area of the
transverse cross section of the tube increases to infinity, the corresponding
sequence of exact $q_c$ or $(q_c)_{\rm eff}$ values is expected to converge to
a limit.  For the tube sections considered here, which have free longitudinal
boundary conditions, this limit is a lower bound for the true $q_c$ of the
infinite simple cubic lattice.  The reason that one can only say that it is a
lower bound is that for these families with free longitudinal boundary
conditions the respective limiting curve ${\cal B}$ exhibits a
complex-conjugate pair of prongs that protrude to the right.  It is possible
that, as the area of the transverse cross section goes to infinity, the
endpoints of these prongs will extend over and meet on the real axis, thereby
defining a point $q_c$ which could lie to the right of the limit of the $q_c$
points for each of the tubes with finite transverse cross section.  In this
case, $\lim_{L_x,L_y \to \infty} q_c(sc, \ (L_x)_{BCx} \times (L_y)_{BCy}
\times \infty_{\rm F})$ is not equal to, but instead less than, the value
$q_c(sc)$ for the infinite simple cubic lattice.  Indeed, for finite-width,
infinite-length strips of the triangular lattice with free longitudinal ($z$)
boundary conditions and periodic transverse $y$ boundary conditions, $\lim_{L_y
\to \infty} q_c((L_y)_{\rm P} \times \infty_{\rm F}) \simeq 3.81967...$ while
$q_c=4$ for the infinite triangular lattice \cite{baxter}.  This occurs because
of the above-mentioned phenomenon in which the endpoints of the
complex-conjugate pair of prongs protruding to the right on the limiting curve
${\cal B}$ move inward toward the real axis and, as $L_y \to \infty$, finally
touch each other, thereby defining the true $q_c=4$ about 5 \% greater than the
above limit at $\sim 3.82$. The presence of these types of prongs protruding to
the right on the limiting curves was also observed for a variety of
infinite-length, finite-width 2D lattice strips with free longitudinal boundary
conditions (for both free and periodic transverse boundary conditions)
\cite{strip,strip2,t,s4,sstransfer,JS}.  Calculations for $L_{\rm P} \times
\infty_{\rm F}$ (i.e. cylindrical) strips of the square lattice have been
carried out in \cite{strip2,s4,sstransfer,JS} with widths $L$ extending up to
13 \cite{JS}, with $q_c(sq, \ 13_{\rm P} \times \infty_{\rm F}) \simeq 2.916$.
This value is within about 3 \% of the value for the infinite square lattice,
$q_c(sq)=3$.  As the width increases, the endpoints of the prongs do move in
toward the real axis.  These calculations for $L_{\rm P} \times \infty_{\rm F}$
strips of the square lattice are consistent with either of the two
possibilities, that $\lim_{L \to \infty} q_c(sq, L_{\rm P} \times \infty_{\rm
F})$ is equal to, or slightly less than, $q_c(sq)$.

It is interesting that the exact calculations of the singular loci ${\cal B}$
for infinite-length, finite-width strips of various lattices with periodic
longitudinal boundary conditions \cite{tpbc} (and either free or periodic
transverse boundary conditions) yielded loci without such complex-conjugate
prongs (or line segments on the real axis) \cite{w,wcy,s4,tk,tor4,t}.  For
these families of strips it was found that ${\cal B}$ always crossed the
positive real axis at $q=0$ and at a maximal point, so $q_c$ is always defined;
furthermore, $L_{\rm F} \times \infty_{\rm P}$, the value of $q_c$ is a
monotonically nondecreasing function of the width.  For $L_{\rm P} \times
\infty_{\rm P}$ strips, although $q_c$ is not a nondecreasing function of the
width, it is, for a given width, closer to the value for the infinite 2D
lattice if one uses periodic, rather than free transverse boundary conditions,
as one would expect since the former minimize finite-size effects.

Some definitions from graph theory will be useful here.  The degree of a vertex
in a graph $G=(V,E)$ is the number of edges to which it is attached.  A graph
$G$ defined to be $\Delta$-regular iff each of its vertices has the same
degree, $\Delta$.  This is, in particular, true of an infinite regular lattice,
where this degree is the coordination number.  Even if some vertices
have different degrees from others, one can define an effective coordination
number,
\beq 
\Delta_{\rm eff} = \lim_{|V| \to \infty} \frac{2|E|}{|V|} \ . 
\label{deltaeff}
\eeq
For example, for a strip of the square lattice of size of the type 
$(L_x)_{\rm F} \times (L_y)_{\rm F}$, in the limit in which the length 
$L_x \to \infty$ with fixed finite width, $L_y$, 
\beq
\Delta_{\rm eff}(sq,(L_x)_{\rm F} \times (L_y)_{\rm F} ;\ L_x \to \infty) = 
4-\frac{2}{L_y} \ . 
\label{deltaff}
\eeq
For the same strip, but with periodic transverse boundary conditions, if $L_y
\ge 3$ to avoid double edges, we have
\beq
\Delta_{\rm eff}(sq,(L_x)_{\rm F} \times (L_y)_{\rm P} \ ;L_x \to \infty) = 4 
\ . 
\label{deltapf}
\eeq
For the $(L_x)_{\rm F} \times (L_y)_{\rm F} \times (L_z)_{\rm F}$ section of 
the simple cubic 
lattice with free boundary conditions in all three directions, the interior 
vertices have 
degree 6, the vertices on the surface away from the corners have degree 4, 
and the corner-vertices have degree 3.  For the minimal case 
$2_{\rm F} \times 2_{\rm F}\times (L_z)_{\rm F}$, there are no interior 
vertices in the 3D sense, so that 
\beq
\Delta_{\rm eff}(sc,2_{\rm F} \times 2_{\rm F} \times (L_z)_{\rm F} ; \ 
L_z \to \infty) = 4 \ . 
\label{delta22z}
\eeq
For the next case $(L_x)_F \times 2_F \times (L_z)_{\rm F}$ 
with $L_x \geq 3$ there are also no interior vertices in the 3D sense; in
this case 
\beq
\Delta_{\rm eff}(sc,(L_x)_{\rm F} \times 2_{\rm F} \times (L_z)_{\rm F};
        FBC_x,FBC_y,FBC_z;L_z \to\infty) = 5 - {2 \over L_x} \ . 
\label{deltax2z}
\eeq
For the $L_z \to \infty$ limit of the $L_x \times L_y \times L_z$ section 
with $L_x \ge 3$, $L_y \ge 3$, we have
\beq
\Delta_{\rm eff}(sc,(L_x)_{\rm F} \times (L_y)_{\rm F} \times (L_z)_{\rm F}; 
\ L_z \to \infty)= 6 - 2\frac{(L_x+L_y)}{L_x L_y} \ . 
\label{deltafff}
\eeq
We shall also use the corresponding formulas when periodic rather than free
boundary conditions are imposed in one of the transverse directions.
Of course, in the usual thermodynamic limit of the Cartesian lattice 
${\mathbb E}^d$, with $L_i \to \infty$, $i=1,..,d$ and 
$\lim_{L_i \to \infty} L_i/L_j$ equal to a finite nonzero constant, the 
effective degree is $\Delta_{\rm eff}({\mathbb E}^d)=2d$ independent of 
boundary conditions. In this case, although free boundary conditions in one 
or more directions entails vertices of lower degree than $2d$, these vertices 
occupy a vanishing fraction of the total vertices in the thermodynamic limit.
Concerning the chromatic number, for the sections of 3D lattices considered
here, we have
\beq
\chi(sc, (L_x)_{\rm F} \times (L_y)_{\rm F} \times (L_z)_{\rm F})=2
\label{chifff}
\eeq
i.e., these graphs are bipartite.  For the sections of the simple cubic 
lattice with periodic boundary conditions in one of the transverse direction,
taking this to be the $x$ direction with no loss of generality, 
\beq
\chi(sc, (L_x)_{\rm P} \times (L_y)_{\rm F} \times (L_z)_{\rm F})=  
                                  \cases{ 2 & if $L_x$ is even \cr
                                          3 & if $L_x$ is odd \cr }
\label{chifpf}
\eeq

For an infinite regular lattice having coordination number $\Delta$, a 
rigorous upper bound on $q_c$ was derived in \cite{ssbounds} using the 
Dobrushin theorem: 
\beq
q_c \le 2\Delta \ . 
\label{qcupper}
\eeq
Aside from the general property that, for cases where there is a $q_c$, this
point satisfies $q_c \ge 2$, we are not aware of a
published lower bound on $q_c$.

We have noted that in the infinite-length limit of given family of lattice
strip graphs, ${\cal B}$ does not necessarily intersect or cross the real $q$
axis, and for such a family, there is no $q_c$.  It has been found that for a
graph consisting of a strip of a regular lattice, a sufficient condition for
${\cal B}$ to cross the (positive) real $q$ axis is that there be periodic
boundary conditions in the longitudinal direction, i.e., the direction in which
the length goes to infinity as $n \to \infty$ \cite{w}-\cite{bcc}.

An interesting question is how $q_c$ depends on the lattice coordination
number, equivalent to the vertex degree for an infinite lattice.  For families
of $L_x \times L_y$ lattice strip graphs with periodic longitudinal ($L_x$)
boundary conditions, whose $L_x \to \infty$ limits are guaranteed to have a
$q_c$, it is found that the value of $q_c$ is a nondecreasing function of the
effective vertex degree $\Delta_{\rm eff}$.  Thus, for the family of cyclic
strips of the square lattice with free transverse boundary conditions, besides
the the $1_F \times \infty_F$ case with $\Delta=2$ and $q_c=2$, one finds the
following results: (i) $2_F \times \infty_F$ ($\Delta=3$) yields $q_c=2$
\cite{w}, (ii) $3_F \times \infty_F$ ($\Delta_{\rm eff}=10/3$) yields $q_c
\simeq 2.3365$ \cite{wcy}, (iii) $3_F \times \infty_F$ ($\Delta_{\rm eff}=7/2$)
yields $q_c \simeq 2.4928$ \cite{s4}.  When one makes both the longitudinal and
transverse boundary conditions periodic, the vertex degree is fixed.  In this
case, $q_c$ does not necessarily increase with increasing width, $L_y$.  For
example, for strips of the square lattice with toroidal boundary conditions and
hence $\Delta=4$, the $3_P \times \infty_P$ strip yields $q_c=3$ \cite{tk}
while the $4_P \times \infty_P$ strips yields the smaller value $q_c \simeq
2.78$ \cite{tor4}.

When one considers strips without periodic longitudinal boundary conditions,
there may or may not be a $q_c$.  We have found that for strips with free
transverse and longitudinal boundary conditions, $q_c$ or $(q_c)_{\rm eff}$,
where the latter can be defined, is a monotonically increasing function of
$L_y$, and since in these families of lattice strips, $\Delta_{\rm eff}$ is a
monotonically increasing function of $L_y$, this means that $(q_c)_{\rm eff}$
is also a monotonically increasing function of $\Delta_{\rm eff}$.  Thus, for
$1_F \times \infty_F$ with $\Delta_{\rm eff}=2$ and $2_F \times \infty_F$ with
$\Delta_{\rm eff}=3$, ${\cal B}=\emptyset$ and there is no $(q_c)_{\rm eff}$,
while $3_F \times \infty_F$ with $\Delta_{\rm eff}=10/3$ yields $q_c=2$
\cite{strip}, $4_F \times \infty_F$ with $\Delta_{\rm eff}=7/2$ yields $q_c
\simeq 2.3014$ \cite{strip}, $5_F \times \infty_F$ with $\Delta_{\rm eff}=3.60$
yields $q_c \simeq 2.428$ \cite{s4,sstransfer}, etc. up to $8_F \times
\infty_F$ with $\Delta_{\rm eff}=3.75$, which yields arc endpoints at $q \simeq
2.6603 \pm 0.0013i$ and hence $(q_c)_{\rm eff}=2.6603$ \cite{sstransfer}.

In contrast, for strips with free longitudinal and 
periodic transverse boundary conditions, $\Delta_{\rm eff}$ is fixed, e.g.,
equal to 4 for strips of the square lattice.  Insofar as one expects $q_c$ or
$(q_c)_{\rm eff}$ to depend on $\Delta_{\rm eff}$, it is not clear, {\it a
priori}, how these quantities should behave as functions of $L_y$.  Indeed, one
finds that $q_c$ or $(q_c)_{\rm eff}$ have no monotonic dependence on $L_y$.
For example, for $3_P \times \infty_F$, ${\cal B}=\emptyset$, $4_P \times
\infty_F$ yields $q_c \simeq 2.3517$ \cite{strip2}, $5_P \times \infty_F$
yields $q_c \simeq 2.6917$ \cite{s4,sstransfer}, but $6_P \times \infty_F$
yields $q_c \simeq 2.6132$ \cite{s4,sstransfer}.  For $7_P \times \infty_F$,
${\cal B}$ has arc endpoints near the real axis at $q \simeq 2.7515 \pm
0.0025i$, so $(q_c)_{\rm eff} \simeq 2.7515$ \cite{sstransfer}.  For the cases
$(L_y)_P \times \infty$ up to $L=11$, it is found empirically that $q_c$ or
$(q_c)_{\rm eff}$ is monotonically increasing separately for the even-$L_y$ and
odd-$L_y$ sequences \cite{JS}.

Since $q_c$ has thermodynamic significance as the value of $q$ above which the
Potts antiferromagnet has no finite-temperature phase transition and is
disordered even at $T=0$, the value of $q_c$ for a lattice with dimensionality
$d \ge 2$ should be independent of the boundary conditions used to take the
thermodynamic limit.  (Strictly speaking, this is not true for $d=1$, since no
$q_c$ is defined for the infinite-length limit of the line graph, while $q_c=2$
for the infinite-length limit of the circuit graph, but this may be considered
to be an exception due to the special nature of this graph.)  For the infinite
2D lattices where $q_c$ is known exactly, it is also a monotonically increasing
function of the lattice coordination number.  Specifically, as noted above, on
the square lattice ($\Delta=4$), one has $q_c=3$ \cite{lieb}.  For the kagom\'e
lattice, again with $\Delta=4$, one finds $q_c=3$ \cite{baxter70}, and on the 
triangular lattice with $\Delta=6$, one has $q_c=4$ \cite{baxter}.

These exact results are all consistent with the inference that the value of
$q_c$ increases as a function of the coordination number for the thermodynamic
limit of a regular lattice or, more generally, for infinite-length strips or
tubes of regular lattices with prescribed boundary conditions in the transverse
directions.  A plausibility argument for this is as follows.  For $q > q_c$,
the zero-temperature Potts antiferromagnet has nonzero ground state entropy per
site, $S = k_B \ln W > 0$, i.e., ground state degeneracy per site $W > 1$.
This entropy reflects the fact that the number of ways of carrying out a proper
coloring of the lattice or lattice strip increases exponentially with the
number of vertices.  Roughly speaking, the constraint that no two adjacent
vertices can be assigned the same color is more restrictive the greater the
number of neighboring vertices there are, i.e., the greater $\Delta_{\rm eff}$
is.  Therefore, as one increases $\Delta_{\rm eff}$ for a fixed $q$, $W$
decreases This is borne out by exact and numerical calculations of $W$ (for 2D
lattices, see, e.g., Fig. 5 of \cite{w} or Fig. 6 of \cite{wn}).  In the
context of $d$-dimensional Cartesian lattices, this means that $q_c$ should
increase as a function of $d$.  Such a monotonicity relation is valuable to
have since the value of $q_c$ is not known exactly for lattices with dimension
$d \ge 3$.

It is useful to define a ratio of the actual value of $q_c$ to the 
upper bound in (\ref{qcupper}), namely, 
\beq
R_{q_c} = \frac{q_c}{2\Delta} \ . 
\label{rqc}
\eeq
Evidently, $R_{q_c}=0.5$ for the infinite-length limit of the circuit graph 
and $R_{q_c}=0.375$ for the infinite square lattice.  Similarly, for other 
exactly known cases, the actual value of $q_c$ is considerably less than the 
rigorous upper bound (\ref{qcupper}).  This suggests that it may be possible 
to improve the upper bound (\ref{qcupper}). 

A generic form for chromatic polynomials for a strip graph of type $G_s$, 
or, more generally, a recursive family of graphs composed of $m$ repetitions 
of a basic subgraph, is \cite{bkw} 
\beq 
P(G_s,m,q)
= \sum_{j=1}^{N_{G_s,\lambda}} c_{G_s,j}(q)(\lambda_{G_s,j}(q))^m
\label{pgsum}
\eeq
where $c_{G_s,j}(q)$ and the $N_{G_s,\lambda}$ terms (eigenvalues) 
$\lambda_{G_s,j}(q)$ depend on the type of
strip graph $G_s$ including the boundary conditions but are independent of 
$m$.  Here, $N_\lambda \le \dim(T)$, where $\dim(T)$ is the 
dimension of the
transfer matrix in the Fortuin-Kasteleyn representation, and the difference, 
$N_{0a}=\dim(T)-N_\lambda$, is the number of zero amplitudes (coefficients). 
The coefficients $c_{G_s,j}$ can be regarded as the multiplicities of the 
eigenvalues $\lambda_{G_s,j}$ (for some real positive $q$ values, these
coefficients can be zero or negative, so that this interpretation presumes a
sufficiently large real positive $q$, followed by analytic continuation to
other values of $q$).

\section{Tubes of the Simple Cubic Lattice with Free Transverse Boundary Conditions}
\label{sec_sc_free}

The computations of the transfer matrix and the chromatic polynomials were
performed following the methods described in Section 3 of Ref.
\cite{sstransfer}.  The idea is to construct the partition function by building
the lattice layer by layer. In all the computations we have chosen the
Fortuin-Kasteleyn representation of the transfer matrix.  Thus, our basis will
be the connectivities of the top layer, whose basis elements ${\bf v}_{\cal P}$
are indexed by partitions ${\cal P}$ of the single-layer vertex set $V^0$.  We
shall abbreviate delta functions $\delta(\sigma_1,\sigma_3)$ as
$\delta_{1,3}$. Moreover, we shall also abbreviate partitions ${\cal P}$ by
writing instead the corresponding product ${\bf v}_{\cal P}$ of delta
functions: e.g., in place of ${\cal P} = \{\{1,3\},\{2,4\},\{5\}\}$ we shall
write simply ${\cal P} = \delta_{1,3}\delta_{2,4}$.  The transfer matrix can be
expressed as
\beq
{\mathsf T} = {\mathsf V} \, {\mathsf H}
\label{def_T}
\eeq
where the matrices ${\mathsf H}$ and ${\mathsf V}$ are defined as
\beqs
{\mathsf H} &=& \prod\limits_{<xx'>\in E^0} \left[ I - J_{xx'} \right] \\
{\mathsf V} &=& \prod\limits_{x\in V^0} \left[- I + D_x \right] \ . 
\label{def_V} 
\eeqs
In these formulae $E^0$ is the single-layer edge set, and 
$J_{x,x'}$ and $D_x$ are respectively 
the ``join'' and ``detach'' operators whose action on the elements 
of the basis ${\bf v}_{\cal P}$ is as follows 
\beqs
J_{xx'} {\bf v}_{\cal P} &=& {\bf v}_{{\cal P}\bullet xx'} \\ 
D_x {\bf v}_{\cal P} &=& \left\{ \begin{array}{ll} 
        {\bf v}_{{\cal P}\setminus x} & {\rm if} \{x\} \notin {\cal P} \\
      q {\bf v}_{\cal P}              & {\rm if} \{x\}    \in {\cal P} 
        \end{array} \right. 
\eeqs
where ${\cal P}\bullet xx'$ is the partition obtained from ${\cal P}$ by
amalgamating the blocks containing $x$ and $x'$ (if they were not already in 
the same block); and ${\cal P}\setminus x$ is the partition obtained from 
${\cal P}$ by detaching $x$ from its block (and thus making it what we term 
a ``singleton'').
In eq. (\ref{def_V}) we have written the formulae for ${\mathsf H}$ and 
${\mathsf V}$ in the zero-temperature Potts antiferromagnet limit; the general 
expressions can be found in \cite{sstransfer}. 

Finally, the partition function can be written as
\beq
Z(L_x\times L_y \times L_z;q) = {\bf u}^{\rm T} \cdot {\mathsf H} \,  
                ({\mathsf V}{\mathsf H})^{L_z-1} \cdot  
                {\bf v}_{\rm id} 
\label{def_Z}
\eeq
where ``id'' denotes the partition in which each site $x\in V^0$ is a 
singleton, and ${\bf u}^{\rm T}$ is defined by
\beq
{\bf u}^{\rm T} \cdot {\bf v}_{\cal P} = q^{|{\cal P}|} \ . 
\eeq

In the zero-temperature limit the horizontal operator ${\mathsf H}$ is
a projection. This implies that we can work in its image subspace by using 
the 
modified transfer matrix ${\mathsf T}' = {\mathsf H}{\mathsf V}{\mathsf H}$ 
in place of ${\mathsf T} = {\mathsf V}{\mathsf H}$, and using the basis vectors
\beq 
{\bf w}_{\cal P} = {\mathsf H}\, {\bf v}_{\cal P} 
\eeq
in place of ${\bf v}_{\cal P}$. Note that ${\bf w}_{\cal P}=0$ if ${\cal P}$
contains nay pair of nearest neighbors in the same block. 
In the following subsections we will list the basis 
${\bf P} = \{{\bf v}_{\cal P}\}$, although we performed the actual
computations with the basis $\{{\bf w}_{\cal P}\}$. 

The computation of the limiting curves ${\cal B}$ was performed with either
the resultant and the direct-search methods (See ref.~\cite{sstransfer}
for details). The computation of the zeros of the chromatic polynomials
was done using the package MPSolve designed by Bini and Fiorentino
\cite{Bini,Bini2}, whose performance is superior to that of 
Mathematica for this particular task \cite{sstransfer}. 

%
%
\subsection{$2_{\rm F} \times 2_{\rm F} \times (L_z)_{\rm F}$ Section of the
Simple Cubic Lattice} \label{sec_sc_2x2F}

Before proceeding to our new results, we note the identity
\beq
sc(2_{\rm F} \times 2_{\rm F} \times (L_z)_{\rm F}) = 
       sq(4_{\rm P} \times (L_z)_{\rm F}) \ . 
\label{2f2flzeq4plz}
\eeq 
The left-hand side of this identity is evidently a tube of the simple
cubic lattice with a minimal-size transverse cross section, viz., a single
square; the right-hand side of the identity is a strip of the square
lattice of width 4 squares, and periodic transverse boundary conditions.  The
chromatic polynomials for the family of graphs on the right-hand side of this
identity were calculated in section 7 of \cite{strip2} and, in the
infinite-length limit, $W$ and ${\cal B}$ were determined (see Fig. 3(a) of
\cite{strip2}).  In Ref. \cite{strip2}, the longitudinal direction 
was taken as $L_x$, while we take it to be $L_z$ here. 
In \cite{strip}-\cite{hs}, a generating function formalism was developed and
used; for a given strip graph $G_s$ of length $m$, the chromatic
polynomial $P((G_s)_m,q)$ is given as the coefficient of $z^m$, where $z$ is
the auxiliary expansion variable:
\beq
\Gamma(G_s,q,z) = \sum_{m=0}^{\infty}P((G_s)_m,q)z^m, 
\label{gamma}
\eeq
This generating function is a rational function of $q$ and $z$.  For the strip 
considered here, of length $L_z=m$ edges, 
\beq
\Gamma(G_s,q,z) = \frac{a_0+a_1z}{1+b_1z+b_2z^2}
\label{gammagen}
\eeq
where
\beq
a_0=q(q-1)(q^2-3q+3)
\label{a0sqtb}
\eeq
\beq
a_1=-q(q-1)(q^4-7q^3+16q^2-13q+5)
\label{a1sqtb}
\eeq
\beq
b_1=-q^4+8q^3-29q^2+55q-46
\label{b1sqtb}
\eeq
\beq
b_2=q^6-12q^5+61q^4-169q^3+269q^2-231q+85
\label{b2sqtb}
\eeq
(Here we have converted the numerator coefficients from \cite{strip2} to be in
accord with the labelling convention of eq. (\ref{gamma}); in the labelling 
convention of \cite{strip2}, the strip with length $m+1$ edges.) 
The chromatic polynomial can equivalently be written as in \ref{def_Z}: in 
the basis ${\bf P} = \{ 1, \delta_{1,4} + \delta_{2,3} \}$ it is given by 
\beq
Z(2_{\rm F} \times 2_{\rm F} \times (L_z)_{\rm F}; q) =  
\left( \begin{array}{c}
             q(q-1)(q^2-3q+3) \\
             2q(q-1)^2
             \end{array}
\right)^{\rm T} \cdot T(2_{\rm F} \times 2_{\rm F})^{L_z -1} \cdot
\left( \begin{array}{c}
             1 \\
             0 \\
            \end{array}
\label{Z_2x2F}
\right)
\eeq
The transfer matrix $T(2_{\rm F} \times 2_{\rm F})$ is equal to  
\beq
T(2_{\rm F}\times 2_{\rm F})   = \left( \begin{array}{cc} 
        41 - 51 q + 28 q^2 - 8 q^3 + q^4 & 2 (-12 + 14 q - 6 q^2 + q^3) \\
        -5 + 2 q                         & 5 - 4q + q^2
             \end{array}
    \right)
\label{T_2x2F}
\eeq
Note that in general, we should have considered an additional element in the
basis, namely the partition $\delta_{1,4}\delta_{3,2}$; however, its
amplitude vanishes identically.  This can be easily understood by noting that
this particular graph (\ref{2f2flzeq4plz}) is planar and
$\delta_{1,4}\delta_{3,2}$ is a crossing partition.

For this family, ${\cal B}$ consists of two complex-conjugate arcs, a
self-conjugate arc that crosses the real axis at $q \simeq 2.3026$, and a line
segment on this axis extending from $q \simeq 2.2534$ to $q \simeq 2.3517$,
which latter point is thus $q_c$.  It should be noted also that although this
graph is included here as the minimal-transverse-size member of a family of
$L_x \times L_y \times L_z$ sections of the simple cubic lattice, it is
degenerate in the sense that it is actually planar, in contrast to the others
that we shall study here.  There are eight endpoints in ${\cal B}$; two of them
are given by the real values $q \simeq 2.3026$ and $q \simeq 2.3517$; the other
six endpoints are the complex-conjugate pairs $q\simeq 0.7098 \pm 2.0427\,i$,
$q\simeq 1.9923\pm 1.5942\,i$, and $q\simeq 2.9953\pm 1.4266\,i$.

%
%
\subsection{$3_{\rm F} \times 2_{\rm F} \times (L_z)_{\rm F}$ Section of 
Simple Cubic Lattice}
\label{sec_sc_3x2F}

\begin{figure}[hbtp]
\centering
\leavevmode
\epsfxsize=2.5in
\begin{center}
\leavevmode
\epsffile{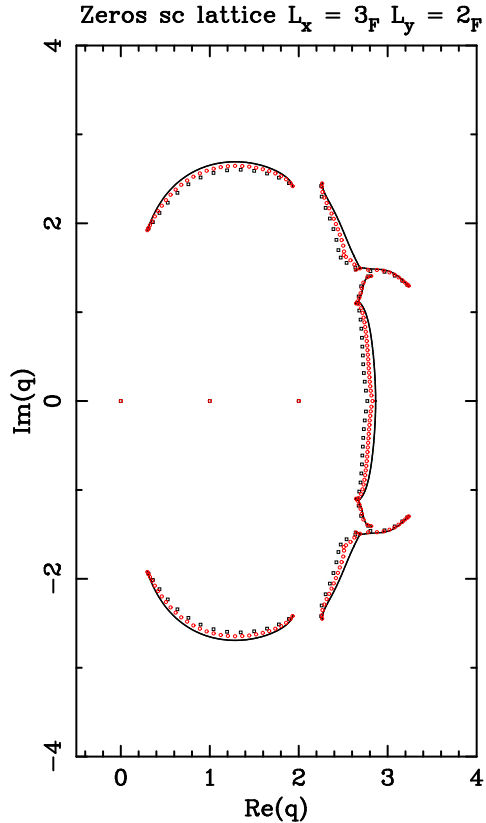}
\end{center}
\caption{\footnotesize{Chromatic zeros for the $3_{\rm F} \times 2_{\rm F} 
\times (L_z)_{\rm F}$
section of the simple cubic lattice, for (a) $L_z=15$, i.e., $n=90$ ($\Box$),
(b) $L_z=30$, i.e., $n=180$ ($\circ$).}}
\label{zeros_sc_3x2F}
\end{figure}

The section of the simple cubic lattice with the next larger transverse cross
section is the $3_{\rm F} \times 2_{\rm F} \times (L_z)_{\rm F}$, or
equivalently, $2_{\rm F} \times 3_{\rm F} \times (L_z)_{\rm F}$, family.  For
this family, the dimension of the transfer matrix is 13.  (In general the
transfer matrix has dimension 15, but two basis elements, namely
$\delta_{1,5}\delta_{2,6}\delta_{3,4} + \delta_{2,4}\delta_{3,5}\delta_{1,6}$
and $\delta_{1,3,5}\delta_{2,4,6}$, can be eliminated as they have zero
amplitudes.)  Since the entries in the transfer matrix are rather long, we
relegate them to the Mathematica file {\tt transfer\_sc.m} that is available
with this paper in the LASL cond-mat archive.  We have computed the chromatic
zeros for $L_z=15$ and 30, i.e., $n=90$ and 180 vertices, respectively.  These
are shown in Fig.~\ref{zeros_sc_3x2F}.  This value of $L_z$ is sufficiently
large that these zeros give a reasonably accurate indication of the location of
the asymptotic limiting curves comprising the locus ${\cal B}$.  As is
generally true, for the larger value of $L_z$, and hence $n$, the chromatic
zeros move slightly outward, approaching the limiting curve ${\cal B}$ from
smaller values of $|q|$.

The limiting curve ${\cal B}$ consists of two complex-conjugate arcs, a
self-conjugate arc that crosses the real $q$-axis at $q\simeq 2.8645$,
and a horizontal real segment from $q\simeq 2.8566$ to $q\simeq 2.8723$.
This latter value corresponds to the value of $q_c$ for this family.
This information is listed, together with that from other lattices, in
Table \ref{qctable}.
There are 16 endpoints: two are the real values $q\simeq 2.8566$ and 
$q\simeq 2.8723$; the other 14 are the complex-conjugated pairs:
$q\simeq 0.2943\pm 1.9150\,i$, $q\simeq 1.9371 \pm  2.4040\,i$,
$q\simeq 2.6374\pm 1.0937\,i$, $q\simeq 2.8250 \pm  1.4087\,i$,
$q\simeq 2.6255\pm 1.4647\,i$, $q\simeq 3.2410 \pm  1.2920\,i$, and
$q\simeq 2.2645\pm 2.4605\,i$. Finally, there are four T points 
at $q\simeq 2.670 \pm  1.117\,i$ and $q\simeq 2.689 \pm 1.500\,i$.  

\bigskip

\begin{table}
\caption{\footnotesize{Results for tube graphs of the simple cubic lattice.
The table shows the relation between the degree (coordination number) $\Delta$
of a $\Delta$-regular family or the effective degree $\Delta_{\rm eff}$, and
$q_c$, if a $q_c$ exists, for the infinite-length limit of the family.
Boundary conditions are indicated with a subscript F for free, P for periodic,
in a given direction.  In column $q_c$, an asterisk indicates that ${\cal B}$
does not actually cross the real axis, so that, strictly speaking, no $q_c$ is
defined, but arcs on ${\cal B}$ end very close to the real axis, at the
positions given.  Similar results are listed for the tubes with $K_{m,m}$ cross
sections.}}
\begin{center}
\begin{tabular}{|c|c|c|c|c|}
\hline\hline $G_s$ & $\Delta$ & $\Delta_{\rm eff}$ & $q_c$ & ref. \\
\hline\hline
sc,  $2_{\rm F} \times 2_{\rm F} \times \infty_{\rm F}$ & $-$ &
     4     & $2.3517$ & \protect\cite{strip2}\\ \hline
sc,  $3_{\rm F} \times 2_{\rm F} \times \infty_{\rm F}$ & $-$ &
     4.33  & $2.8723$ & here \\ \hline
sc,  $4_{\rm F} \times 2_{\rm F} \times \infty_{\rm F}$ & $-$ &
     4.50  & $3.1498 \pm 0.0021^\star$ & here \\
\hline\hline
sc,  $2_{\rm P} \times 2_{\rm P} \times \infty_{\rm F}$ & $-$ &
     4     & $2.3517$ & \protect\cite{strip2}  \\ \hline
sc,  $3_{\rm P} \times 2_{\rm P} \times \infty_{\rm F}$ & $-$ & 5 &
     $3.3255 \pm  0.0184\,i^*$ & here   \\ \hline
sc,  $4_{\rm P} \times 2_{\rm P} \times \infty_{\rm F}$ & $-$ & 5 &
     $3.3623 \pm  0.0061\,i^*$ & here   \\
\hline\hline
$K_{2,2} \times \infty_{\rm F}$ & $-$ & 4 & $2.3517$ &
                \protect\cite{strip2} \\ \hline
$K_{3,3} \times \infty_{\rm F}$ & $-$ & 5 & $3.0452 \pm 0.0082\,i^*$ &
                here \\ \hline
$K_{4,4} \times \infty_{\rm F}$ & $-$ & 6 & $3.6743 \pm 0.0085\,i^*$ &
                here \\
\hline\hline
\end{tabular}
\end{center}
\label{qctable}
\end{table}

%
%
\subsection{$4_{\rm F} \times 2_{\rm F} \times (L_z)_{\rm F}$ Section of 
Simple Cubic Lattice} \label{sec_sc_4x2F}

\begin{figure}[hbtp]
\centering
\leavevmode
\epsfxsize=2.5in
\begin{center}
\leavevmode
\epsffile{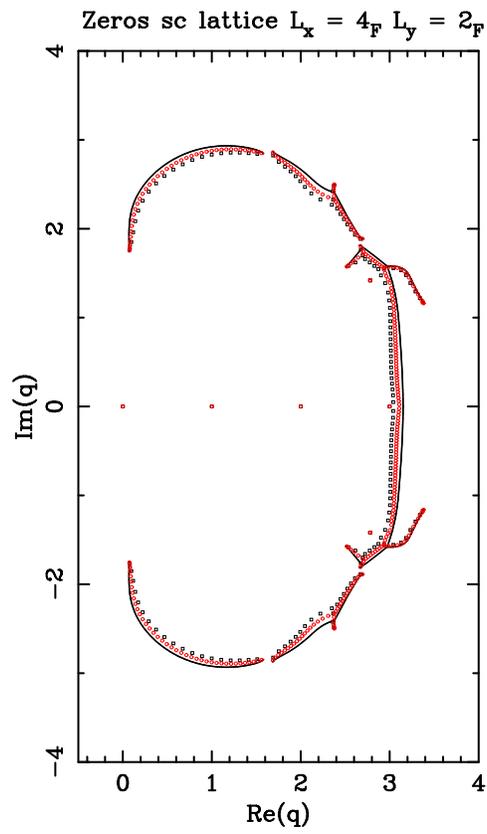}
\end{center}
\caption{\footnotesize{Chromatic zeros for the $4_{\rm F} \times 2_{\rm F} 
\times (L_z)_{\rm F}$
section of the simple cubic lattice, for (a) $L_z=20$, i.e., $n=160$ ($\Box$),
(b) $L_z=40$, i.e., $n=320$ ($\circ$).}}
\label{zeros_sc_4x2F}
\end{figure}

The section of the simple cubic lattice with the next larger transverse cross
section is the $4_{\rm F} \times 2_{\rm F} \times (L_z)_{\rm F}$, 
or equivalently, $2_{\rm F} \times
4_{\rm F} \times (L_z)_{\rm F}$, family.  For this family, the dimension 
of the transfer matrix is 156. However, there are 20 basis elements that
correspond to zero amplitudes, so the effective dimension of the transfer
matrix is 136. This transfer matrix $T(4_{\rm F}\times2_{\rm F})$, 
as well as the vectors ${\bf v}$ and ${\bf u}_{\rm id}$ can be found in 
the {\sc Mathematica} file {\tt transfer\_sc.m}.  
We have computed the chromatic zeros for $L_z=20$ and 40,
i.e., $n=160$ and 320 vertices, respectively.  These are shown in
Fig.~\ref{zeros_sc_4x2F}.  

The limiting curve ${\cal B}$ consists of eight complex-conjugate arcs;
none of them crosses the real axis, so, strictly speaking, there is
no value of $q_c$. However, we can define $(q_c)_{\rm eff}\approx 3.1498$.
This value is larger than the corresponding value for the family
$3_{\rm F}\times 2_{\rm F} \times L_{\rm F}$ $q_c\approx 2.8723$ and larger
than the value for the square lattice $q_c(sq)=3$.
There are 11 pairs of complex-conjugate endpoints:
$q\simeq 0.073 \pm 1.747\,i$,
$q\simeq 1.582 \pm 2.845\,i$, $q\simeq 1.684 \pm 2.871\,i$,
$q\simeq 2.379 \pm 2.503\,i$,
$q\simeq 2.513 \pm 1.569\,i$,
$q\simeq 2.661 \pm 1.816\,i$, $q\simeq 2.697 \pm 1.889\,i$,
$q\simeq 3.034 \pm 1.367\,i$, $q\simeq 3.035 \pm 1.385\,i$,
$q\simeq 3.150 \pm 0.002\,i$, and
$q\simeq 3.383 \pm 1.157\,i$.
Finally, there are three pairs of complex-conjugate T points:
$q\simeq 2.381 \pm 2.404\,i$,
$q\simeq 2.695 \pm 1.793\,i$, and
$q\simeq 2.972 \pm 1.578\,i$.

\section{Tubes of the Simple Cubic Lattice with Transverse Periodic Boundary 
Conditions}
\label{sec_sc_cyl}

The method we have used to compute the chromatic polynomials and the
transfer matrix is the same as in Section~\ref{sec_sc_free}. The only
difference is that we enlarge the single-layer edge set $E^0$.

%
%
\subsection{$2_{\rm P} \times 2_{\rm P} \times (L_z)_{\rm F}$ Section of
Simple Cubic Lattice} \label{sec_sc_2x2P}

This family has (trivially) the same chromatic polynomials as the family 
$sc(2_{\rm F} \times 2_{\rm F} \times (L_z)_{\rm F}$.  (This is not true for
the finite-temperature Potts model partition function, or equivalently, the
Tutte polynomial; however we only deal with the chromatic polynomial here.) 
Thus, the transfer matrix and chromatic polynomials are given 
respectively by \reff{def_T} and \reff{def_Z}.   

%
%
\subsection{$3_{\rm P} \times 2_{\rm P} \times (L_z)_{\rm F}$ Section of 
Simple Cubic Lattice} \label{sec_sc_3x2P}

In general, finite-size artifacts are minimized if one uses periodic boundary
conditions in as many directions as possible.  Hence, it is useful to calculate
$P$ and ${\cal B}$ for the same section of the simple cubic lattice as in the
previous section, but with periodic boundary conditions imposed on the longer
of the two transverse directions.  This family is again bipartite.  In this
case, the dimension of the transfer matrix is 4.  

In the basis ${\bf P} = \{ 1, \delta_{2,4} + \delta_{2,6} + \delta_{1,5} + 
\delta_{3,5} + \delta_{1,6} + \delta_{3,4}, \delta_{1,5}\delta_{2,6} + 
\delta_{2,4}\delta_{3,5} + \delta_{3,4}\delta_{1,5} + \delta_{3,4}\delta_{2,6}
+ \delta_{1,6}\delta_{3,5} + \delta_{1,6}\delta_{2,4}, \delta_{1,6}\delta_{3,4}
+ \delta_{1,5}\delta_{2,4} + \delta_{2,6}\delta_{3,5}\}$, the transfer matrix
can be written as 
\beq
T(3_{\rm P} \times 2_{\rm P}) = \left( \begin{array}{cccc}
    T_{11} & T_{12} & T_{13} & T_{14} \\
    T_{21} & T_{22} & T_{23} & T_{24} \\
    1 & 
    2 (-3 + q) &  
   12 - 6 q + q^2 & 
   -2 (-4 + q) \\
   0 & 
   -2 & 
   -4 (-3 + q) & 
   15 - 7 q + q^2
   \end{array}
   \right)
\eeq
where
\beqs
T_{11} &=& 1089 - 1578 q + 1054 q^2 - 417 q^3 + 103 q^4 - 15 q^5 + q^6 \\ 
T_{12} &=& 6 (-292 + 394 q - 231 q^2 + 74 q^3 - 13 q^4 + q^5) \\
T_{13} &=& 6 (110 - 117 q + 51 q^2 - 11 q^3 + q^4) \\
T_{14} &=& 3 (154 - 149 q + 60 q^2 - 12 q^3 + q^4) \\
T_{21} &=& -90 + 71 q - 20 q^2 + 2 q^3 \\
T_{22} &=& 203 - 189 q + 71 q^2 - 13 q^3 + q^4 \\
T_{23} &=& -109 + 80 q - 21 q^2 + 2 q^3 \\
T_{24} &=& -81 + 52 q - 12 q^2 + q^3
\eeqs
The vectors ${\bf v}$ and ${\bf u}_{\rm id}$ are given by 
\beqs 
{\bf v} &=& \left( \begin{array}{c} 
                   (-2 + q) (-1 + q) q (-13 + 14 q - 6 q^2 + q^3) \\
                   6 (-2 + q)^3 (-1 + q) q  \\
                   6 (-2 + q)^2 (-1 + q) q \\
                   3 (-3 + q) (-2 + q) (-1 + q) q 
                   \end{array}
            \right) \\
{\bf u}_{\rm id} &=& \left( \begin{array}{c}
                    1 \\ 0 \\ 0 \\ 0 
                   \end{array}
            \right)
\eeqs
We remark that in this case the most general basis contains an additional 
partition: $\delta_{1,5}\delta_{2,6}\delta_{3,4} + 
\delta_{2,4}\delta_{3,5}\delta_{1,6}$. This one can be dropped, as it
corresponds to a vanishing amplitude. 

\begin{figure}[hbtp]
\centering
\leavevmode
\epsfxsize=2.5in
\begin{center}
\leavevmode
\epsffile{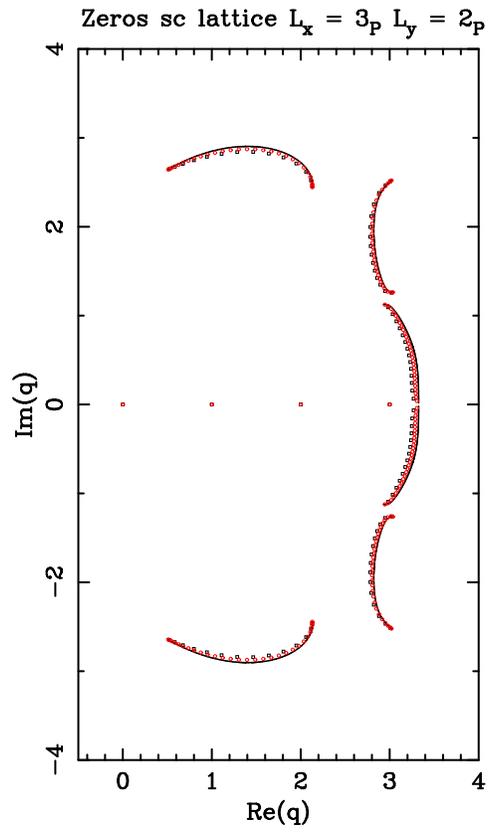}
\end{center}
\caption{\footnotesize{Chromatic zeros for the 
$2_{\rm P} \times 3_{\rm P} \times (L_z)_{\rm F}$
section of the simple cubic lattice, for (a) $L_z=15$, i.e., $n=90$ ($\Box$),
(b) $L_z=30$, i.e., $n=180$ ($\circ$).}}
\label{zeros_sc_3x2P}
\end{figure}

We have computed the chromatic zeros for $L_z=15$ and 30, i.e., $n=90$ and
180, respectively.  These are shown in Fig.~\ref{zeros_sc_3x2P}.
The limiting curve ${\cal B}$ contains three pairs of self-conjugated arcs. 
Although the arcs do not cross the real axis, so that, strictly speaking, no
$q_c$ is defined, $(q_c)_{\rm eff} \simeq 3.33$, which is larger than 
the value $q_c=3$ for the square lattice.  The locus ${\cal B}$ has 
12 endpoints: $q\simeq 0.5061 \pm 2.6413\,i$, 
$q\simeq 2.1301 \pm  2.4407\,i$, $q\simeq 3.0251 \pm 2.5249\,i$, 
$q\simeq 3.0412 \pm  1.2643\,i$, $q\simeq 2.9328 \pm 1.1238\,i$, and  
$q\simeq 3.3255 \pm  0.01839\,i$. 

%
%
\subsection{$4_{\rm P} \times 2_{\rm P} \times (L_z)_{\rm F}$ Section of
Simple Cubic Lattice} \label{sec_sc_4x2P}

In this case we have 56 basis elements. However, there are 11 trivial basis
element that lead to a vanishing amplitude.  The transfer matrix and the
vectors ${\bf v}$ and ${\bf u}_{\rm id}$ are listed in the {\sc Mathematica}
file {\tt transfer\_sc.m}.  Among the other 45 basis elements, we have
numerical indications that there are 19 additional vanishing amplitudes.

\begin{figure}[hbtp]\centering
\leavevmode\epsfxsize=2.5in
\begin{center}\leavevmode
\epsffile{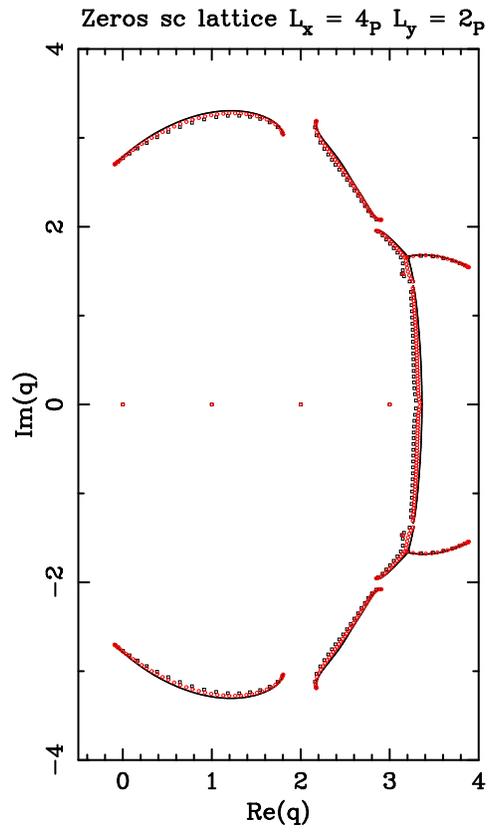}
\end{center}
\caption{\footnotesize{Chromatic zeros for the
$2_{\rm P} \times 4_{\rm P} \times (L_z)_{\rm F}$
section of the simple cubic lattice, for (a) $L_z=20$, i.e., $n=160$ ($\Box$),
(b) $L_z=40$, i.e., $n=320$ ($\circ$).}}
\label{zeros_sc_4x2P}
\end{figure}

The limiting curve ${\cal B}$ (see Fig.~\ref{zeros_sc_4x2P}) 
contains eight self-conjugate arcs. As in the
$4_{\rm P} \times 2_{\rm P} \times \infty_{\rm F}$ case, this locus does not
actually cross the positive real $q$ axis, but again has endpoints that lie
very close to this axis, and we obtain $(q_c)_{\rm eff} \simeq 3.36$.
As expected, this is larger than the value $(q_c)_{\rm eff} \simeq 3.33$ that
we found for the tube $3_{\rm P} \times 2_{\rm P} \times \infty_{\rm F}$.
There are nine pairs of complex-conjugate endpoints:
$q\simeq  -0.095 \pm 2.700 \,i$,
$q\simeq   1.803 \pm 3.031 \,i$,
$q\simeq   2.176 \pm 3.196 \,i$,
$q\simeq   2.841 \pm 1.954 \,i$,
$q\simeq   2.910 \pm 2.080 \,i$,
$q\simeq   3.263 \pm 1.333 \,i$,
$q\simeq   3.279 \pm 1.377 \,i$,
$q\simeq   3.362 \pm 0.006 \,i$, and
$q\simeq   3.892 \pm 1.542 \,i$.
Finally, there is one pair of complex-conjugate T points:
$q\simeq   3.204 \pm 1.665 \,i$.  

Another property that was observed in our earlier calculations
of chromatic polynomials for strips of 2D lattices is that, for a given type
of transverse boundary conditions, for infinite-length strips with free
longitudinal boundary conditions, as the width increases, the number of
arcs on ${\cal B}$ increases and the endpoints of these arcs move in such a
manner as to reduce the gaps between the arcs.  Here we observe the
same qualitative behavior as the area of the transverse cross section of the
tube section increases.  If this trend
continues for progressively larger transverse widths or cross
sectional areas, then the number of arcs could increase without bound as one
approaches the respective infinite 2D or 3D lattices.  Now in
the two cases where the loci ${\cal B}$ have been calculated exactly for
regular lattices, namely the $d=1$ lattice (with periodic boundary conditions;
for free boundary conditions, ${\cal B}=\emptyset$) and the triangular lattice
(defined as the limit $L \to \infty$ of $L_{\rm P} \times \infty_F$ strips
\cite{baxter}), this locus has no prongs or endpoints.  Thus, one could imagine
that as the width or cross sectional area of the strips or tubes increases to
infinity, the endpoints of arcs join so that the gaps between these arcs
disappear, and prongs either have lengths that go to zero or have endpoints
that join to form closed boundaries.  We also observe that, for a given
width or, for the present families of tube sections, for a given transverse
cross sectional area, there tend to be fewer arcs when one uses periodic
rather than free boundary conditions for the transverse direction(s).  This
is in accord with the fact that calculations for strips with periodic
longitudinal boundary conditions \cite{w,wcy,tk,s4,tor4,t} found no prongs
(or line segments) on the respective loci ${\cal B}$, i.e. in all cases
studied, these loci did not contain endpoints.  Finally, our calculations are
consistent with the expectation that as the area of the transverse cross
section goes to infinity, the outer envelope of the locus ${\cal B}$ approaches
a limit, which crosses the real axis at $q=0$, $q_c(sc)$, and other point(s)
between these two.

\section{Further Remarks on the Structure of ${\cal B}$}

In the introductory section, we discussed some differences in the locus ${\cal
B}$ that depend on whether one uses periodic or free longitudinal boundary
conditions.  Here we include some further remarks relevant to our present
results.  It has been established that the structure of ${\cal B}$ inside of
its outer envelope and, in particular, the question of whether and where it
crosses the positive real axis between $q=0$ and $q=q_c(\{G\})$ are dependent
upon the boundary conditions used.  For example, exact calculations for $L_{\rm
F} \times \infty_{\rm P}$ and $L_{\rm P} \times \infty_{\rm P}$ strips of the
triangular lattice show that the respective loci ${\cal B}$ cross the real axis
at $q=2$ \cite{wcy,t,tor4}, corresponding to the fact that the Ising
antiferromagnet has a (frustrated) $T=0$ critical point on these strips;
however the loci ${\cal B}$ obtained in \cite{baxter,strip2} for $L_{\rm P}
\times \infty_{\rm F}$ strips or in \cite{strip} for $L_{\rm F} \times
\infty_{\rm F}$ strips of the triangular lattice do not, in general, cross the
real axis at $q=2$, nor is this crossing obtained for the infinite-width limit
of the cylindrical strips calculated in \cite{baxter}.  Similarly, for $L_{\rm
F} \times \infty_{\rm P}$ and $L_{\rm P} \times \infty_{\rm P}$ strips of the
square lattice it was found that ${\cal B}$ passes through $q=0$, $q=2$, and a
maximal value, $q=q_c(\{G\})$ \cite{w,wcy,tk,s4}. In contrast, for the
corresponding strips of the square lattice with free longitudinal boundary
conditions, it has been found that ${\cal B}$ does not contain $q=0$ or, in
general, $q=2$, and while the arc endpoints nearest to the origin move toward
this point as the width increases (leading to the inference that for infinite
width, ${\cal B}$ would pass through $q=0$), there is no analogous tendency of
arcs on ${\cal B}$ to elongate toward $q=2$ \cite{strip,strip2,sstransfer,JS}.
There is also no tendency for the arcs on the loci ${\cal B}$ for our tube
sections of the simple cubic lattice with free longitudinal boundary conditions
to move toward $q=2$ as the cross sectional area increases.

A second point concerns the positions of the leftmost arcs.  We find from the
exact calculations reported here that for the infinite-length tube sections of
the simple cubic lattice of the form $(L_x)_{\rm P} \times (L_y)_{\rm P} \times
\infty_F$ with sufficiently large transverse cross section, namely $L_x=4$,
$L_y=2$, ${\cal B}$ contains support in the $Re(q) < 0$ half-plane.  This 
suggests that for this family of tube sections of the simple cubic lattice with
periodic transverse boundary conditions, as the transverse area $L_xL_y \to
\infty$, the complex-conjugate curves on ${\cal B}$ could approach the origin
from the $Re(q) < 0$ half-plane, as happens for the limit $L \to \infty$ of
$L_{\rm P} \times \infty_{\rm F}$ strips of the triangular lattice
\cite{baxter} and sufficiently wide strips of the square lattice of the form 
$L_{\rm F} \times \infty_{\rm F}$ \cite{strip,sstransfer}, 
$L_{\rm P} \times \infty_{\rm F}$ \cite{s4,sstransfer,JS}, and 
$L_{\rm F} \times \infty_{\rm P}$ \cite{wcy,pm,s4}.

\section{Behavior of $q_c$ for ${\mathbb E}^d$ with large $d$}

For the $d$-dimensional Cartesian lattice ${\mathbb E}^d$, Mattis suggested 
the ansatz \cite{mattis}
\beq
W({\mathbb E}^n,q) \simeq 1 + \frac{(q-2)^d}{(q-1)^{d-1}} \ . 
\label{wmattis}
\eeq
This agrees with the general results $W(\{C\},q)=q-1$ for $d=1$, and, for the
square lattice ${\mathbb E}^2$ with $q=3$ yields the estimate $W(sq,q=3)=3/2$,
which is within 4 \% of the known result $W({\mathbb
E}^2,q=3)=(4/3)^{3/2}=1.5396..$ \cite{lieb}.  Mattis addressed the question of
$d_c(q)$, i.e. the lower critical dimensionality of the $q$-state Potts
antiferromagnet, below which it is disordered for $T \ge 0$.  An equation
yielding $d_c$ as a function of $q$ can also be solved to yield $q_c$ as a
function of $d$, so $d_c(q)$ and $q_c(d)$ constitute equivalent information
about the system.  Mattis argued that $d_c(q)$ could be estimated by noting
that $W$ is a measure of disorder, and if it is significantly greater than 1,
then the system would be sufficiently disordered that one would not expect
there to be a phase transition at $T \ge 0$. Taking the criterion that $W < 2$
as the demarcation value for which a zero-temperature phase transition could
occur, this yields the result
\beq 
d_c(q) = \frac{\ln(q-1)}{\ln \Bigl ( \frac{(q-1)}{(q-2)} \Bigr )} \ . 
\label{dc}
\eeq
As noted, this may equivalently be regarded as an equation for $q_c$ as a 
function of $d$ and for $d=3$, this ansatz gives 
\beq
q_c(d=3) \simeq 4.15
\label{qcd3}
\eeq
Our exact results $W$ and $q_c$ for tube sections of the simple cubic are 
consistent with this estimate from the ansatz (\ref{wmattis}). 

It is also of interest to ask what the behavior of $q_c$ is for large $d$ on
Cartesian lattices.  The ansatz (\ref{wmattis}) yields the asymptotic behavior
\beq
q_c \sim \frac{d}{\ln d} \quad {\rm for} \ \ d \to \infty  \ . 
\label{qcasymptotic}
\eeq
This is in agreement with the upper bound (\ref{qcupper}) 
\beq
q_c \le 4d
\label{qcupperd}
\eeq
and evidently is a smaller and smaller fraction of this upper bound as $d$ gets
large, with 
\beq
R_{q_c} \sim \frac{1}{4 \ln d} \quad {\rm for} \ \ d \to \infty \ . 
\label{rupperd}
\eeq
 
\section{Family $(K_{m,m})^{L_z}$}
\label{sec_Knn}

It is also useful to calculate ${\cal B}$ and study the dependence of $q_c$ on
vertex degree for infinite-length limits of other families of tube graphs.  We
report calculations here for a family of tube graphs whose transverse cross
section is the complete bipartite graph $K_{m,m}$. The graph $K_{m,m}$ is
$\Delta$-regular graphs with $\Delta=m$.  We construct our tubes with the
$K_{m,m}$ transverse cross section connected lengthwise $L_z$ times, so that
each vertex of one $K_{m,m}$ subgraph is connected ``vertically'' to the
corresponding vertex of the next $K_{m,m}$.  This recursive family of graphs is
denoted as $(K_{m,m})^{L_z}$. The value of $\Delta_{\rm eff}$ is for this
family
\beq
\Delta_{\rm eff}\left( (K_{m,m})^{L_z}; FBC_z; L_z \rightarrow\infty \right) 
  = m + 2 \ . 
\eeq
The computational method is the same as in the simple-cubic families: the 
transfer matrix and the partition function are computed using the same
formulae \reff{def_T}/\reff{def_Z}. The only difference is the single-layer
edge set $E^0$. 

%
%
\subsection{Family $(K_{2,2})^{L_z}$}
\label{sec_K22}

The family $(K_{2,2})^{L_z}$ is trivially equivalent to the family ${\rm
sq}(4_{\rm P}\times L_{\rm F})$, so by \reff{2f2flzeq4plz} it is also
equivalent to the family ${\rm sc}(2_{\rm F}\times 2_{\rm F} \times 
(L_z)_{\rm F})$
The transfer matrix, written in the basis ${\bf P} = \{ 1, \delta_{1,2} +
\delta_{3,4} \}$, takes the same form as \reff{T_2x2F}; and the partition
function is equivalent to \reff{Z_2x2F}.

%
%
\subsection{Family $(K_{3,3})^{L_z}$}
\label{sec_K33}

The transfer matrix for the family $(K_{3,3})^{L_z}$ has dimension 5. In the 
basis ${\bf P} = \{ 1, \delta_{1,2} + \delta_{1,3} + \delta_{2,3} + 
\delta_{4,5} + \delta_{4,6} + \delta_{5,6}, \delta_{1,2,3} + \delta_{4,5,6}, 
\delta_{1,2}\delta_{4,5} + \delta_{1,2}\delta_{4,6} +\delta_{1,2}\delta_{5,6}+
\delta_{1,3}\delta_{4,5} + \delta_{1,3}\delta_{4,6} +\delta_{1,3}\delta_{5,6}+ 
\delta_{2,3}\delta_{4,5} + \delta_{2,3}\delta_{4,6} +\delta_{2,3}\delta_{5,6}, 
\delta_{1,2,3}\delta_{4,5} + \delta_{1,2,3}\delta_{4,6} + 
\delta_{1,2,3}\delta_{5,6} + \delta_{4,5,6} \delta_{1,2} + 
\delta_{4,5,6}\delta_{1,3} + \delta_{4,5,6}\delta_{2,3} \}$,   
we can write the transfer matrix as 

\beq
T(K_{33}) = \left( \begin{array}{ccccc}
  T_{11} & T_{12} & T_{13} & T_{14} & T_{15} \\ 
  T_{21} & T_{22} & T_{23} & T_{24} & T_{25} \\ 
  T_{31} & T_{32} & T_{33} & T_{34} & T_{35} \\ 
  1      & 2(-3 + q) & 0 & T_{44} & 2(-2 + q)\\ 
  0      & 1      & 0      & -3     & 2-q 
  \end{array}\right) 
\eeq
where
\beqs
T_{11} &=& 1234 - 1747 q + 1137 q^2 - 437 q^3 + 105 q^4 - 15 q^5 + q^6 \\ 
T_{12} &=& 6 (-252 + 337 q - 198 q^2 + 65 q^3 - 12 q^4 + q^5) \\
T_{13} &=& 2 (36 - 56 q + 33 q^2 - 9 q^3 + q^4) \\ 
T_{14} &=& 9 (89 - 94 q + 43 q^2 - 10 q^3 + q^4) \\ 
T_{15} &=& 6 (-25 + 24 q - 8 q^2 + q^3) \\
T_{21} &=& -113 + 91 q - 27 q^2 + 3 q^3 \\
T_{22} &=& 174 - 153 q + 54 q^2 - 10 q^3 + q^4 \\ 
T_{23} &=& -10 + 13 q - 6 q^2 + q^3 \\ 
T_{24} &=& 3 (-40 + 28 q - 8 q^2 + q^3) \\ 
T_{25} &=& 29 - 21 q + 4 q^2 \\ 
T_{31} &=& 55 - 27 q + 3 q^2 \\
T_{32} &=& -3(29 - 17q + 3q^2) \\
T_{33} &=& 9 - 12q + 6q^2 - q^3\\
T_{34} &=& -18(-3 + q) \\
T_{35} &=& -3(7 - 5q + q^2)\\
T_{44} &=&12 - 5q + q^2
\eeqs
Finally, 
\beqs 
{\bf v} &=& \left( \begin{array}{c}
(-1 + q) q (31 - 47 q + 28 q^2 - 8 q^3 + q^4)\\
   6 (-1 + q) q (-7 + 10 q - 5 q^2 + q^3)\\
   2 (-1 + q)^3 q \\ 
   9 (-1 + q) q (3 - 3 q + q^2) \\ 
   6 (-1 + q)^2 q
                   \end{array} \right) \\
{\bf u}_{\rm id} &=& \left( \begin{array}{c}
   1 \\ 0 \\ 0 \\ 0 \\ 0 
                   \end{array} \right)
\eeqs

\bigskip
We remark that in this case there is an additional element of the basis 
$\delta_{1,2,3}\delta_{4,5,6}$ that should be taken into account in general.
However, it corresponds to a vanishing amplitude.

\bigskip  

\begin{figure}[hbtp]
\centering
\leavevmode
\epsfxsize=2.5in
\begin{center}
\leavevmode
\epsffile{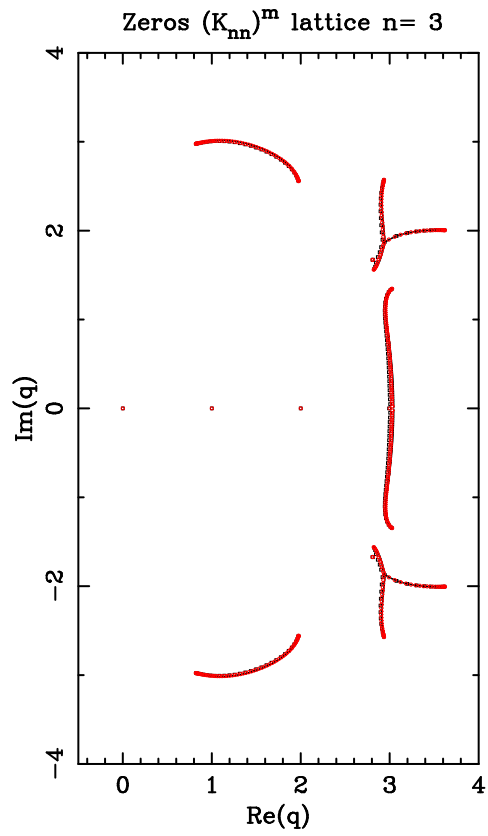}
\end{center}
\caption{\footnotesize{Chromatic zeros for the $(K_{3,3})^{m}$ graph for (a)
$m=L_z=15$, i.e., $n=90$ ($\Box$), (b) $m=L_z=30$, i.e. $n=180$ ($\circ$).}}
\label{zeros_K_33}
\end{figure}

Chromatic zeros for $(K_{3,3})^{L_z}$ with $L_z=15$ and $L_z=30$ are shown in 
Fig.~\ref{zeros_K_33}, as well as the limiting curve ${\cal B}$. 
The limiting curve ${\cal B}$ contains six connected pieces. None of them 
crosses the real axis. Thus, strictly speaking, there is no $q_c$ defined.
However, by extrapolating the closest points to the real axis we get 
$q_c \simeq 3.045$, which is slightly greater than the value for the square
lattice $q_c=3$.

There are 14 endpoints: $q\simeq 0.8197 \pm 2.9764\,i$, 
$q\simeq 1.9761 \pm 2.5559\,i$, 
$q \simeq 2.8190 \pm 1.5587\,i$, $q \simeq 2.9364 \pm 2.5742\,i$, 
$q \simeq 3.6220 \pm 2.0051\,i$, $q \simeq 3.0283 \pm 1.3476\,i$, 
and $q\simeq 3.0452 \pm 0.008246\,i$. There are two complex-conjugated 
T points at $q\simeq 2.949 \pm 1.870\,i$. 

%
%
\subsection{Family $(K_{4,4})^{L_z}$}
\label{sec_K44}

In this case the transfer matrix has 15 elements. However, three of them 
correspond to null amplitudes, so we have an effective 12-dimensional 
transfer matrix. This matrix is listed 
in the {\sc Mathematica} file {\tt transfer\_Knn\_tube.m} that is available 
with this paper in the LASL cond-mat archive.  

\begin{figure}[hbtp]
\centering
\leavevmode
\epsfxsize=2.5in
\begin{center}
\leavevmode
\epsffile{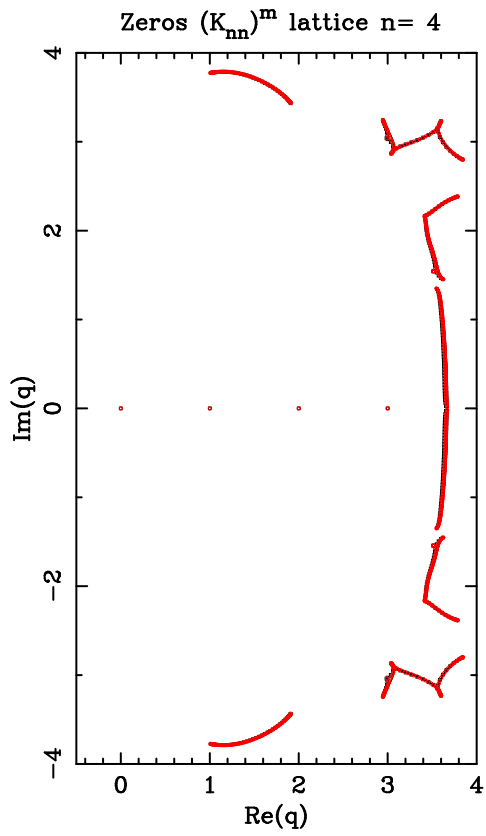}
\end{center}
\caption{\footnotesize{Chromatic zeros for the $(K_{4,4})^{m}$ graph for (a)
$m=L_z=40$, i.e., $n=320$ ($\Box$), (b) $m=L_z=80$, i.e. $n=640$ ($\circ$).}}
\label{zeros_K_44}
\end{figure}

There are eight connected pieces (See Figure~\ref{zeros_K_44}), and none of
them crosses the real axis. The closest points to that axis are the 
complex-conjugated pair $q\approx 3.6743 \pm 0.0085\,i$.  
There are ten endpoints (that were computed using the resultant method): 
$q\approx 1.0084 \pm 3.7740\,i$, $q\approx 1.9104 \pm 3.4341\,i$,  
$q\approx 2.9457 \pm 3.2436\,i$, $q\approx 3.0385 \pm 2.8658\,i$,
$q\approx 3.5456 \pm 1.3512\,i$, $q\approx 3.6006 \pm 3.2332\,i$, 
$q\approx 3.6260 \pm 1.4516\,i$, $q\approx 3.6743 \pm 0.0085\,i$, 
$q\approx 3.7857 \pm 2.3839\,i$, and $q\approx 3.8460 \pm 2.7980\,i$.  
There are four T-points at $q\approx 3.070\pm 2.904\,i$, and 
$q\approx 3.567\pm 3.158\,i$. 

\section{Conclusions}

In this paper we have reported exact solutions for the zero-temperature
partition function of the $q$-state Potts antiferromagnet on tubes of the
simple cubic lattice with various transverse cross sections and boundary
conditions and with arbitrarily great length.  We have used these to calculate,
in the infinite-length limit, the resultant ground state degeneracy per site
$W$ and the singular locus ${\cal B}$ which is the continuous accumulation set
of the chromatic zeros.  In particular, we have calculated the value of $q_c$
or $(q_c)_{\rm eff}$ for these infinite-length tubes.  Our results show
quantitatively how this quantity increases as the effective coordination number
for a given family of graphs increases and are a step toward determining 
$q_c$ is for the infinite simple cubic lattice.  We have also presented similar
calculations for another interesting family of tube graphs whose transverse 
cross section is formed from the complete bipartite graph $K_{m,m}$.

Acknowledgment:  The research of R. S. was supported in part by the NSF grant
PHY-9722101 The research of J.S. was partially supported by CICyT (Spain)
grants AEN97-1880 and AEN99-0990. J.S. would like to acknowledge the warm
hospitality of the C.N.~Yang Institute for Theoretical Physics, where this work
was initiated. We thank S.-C. Chang and A. Sokal for recent discussions on
related research projects.

\vfill
\eject

\begin{thebibliography}{99}
\bibitem{qcni}{We comment here on cases where $q_c(\{G\})$ is non-integral.
The Potts antiferromagnet can formally be defined for positive real, as well as
integral, $q$, via the formula $Z(G,q,v)=\sum_{G^\prime \subseteq G}
q^{k(G^\prime)}v^{e(G^\prime)}$, where $G^\prime$ is a spanning subgraph of $G$
(i.e. a subgraph containing all of the vertices and a subset of the edges of
$G$), and $k(G^\prime)$ and $e(G^\prime)$ denote the number of connected
components and edges in $G^\prime$.  However, since $-1 \le v \le 0$ for the
Potts antiferromagnet, this model does not, in general, have a Gibbs measure
for non-integral $q$, and this can lead to unphysical behavior
\cite{ssbounds,a} such as negative specific heat and non-existence of a
thermodynamic limit independent of boundary conditions. In contrast, for
positive integral $q$, one can always define the partition function for the
Potts antiferromagnet as $Z(G,q,v)=\sum_{\sigma_i}\exp(-\beta {\cal H})$ where
$\beta=(k_BT)^{-1}$ and ${\cal H}=-J\sum_{\langle ij \rangle}
\delta_{\sigma_i,\sigma_j}$ with $J < 0$, which obviously has a Gibbs
measure.}
\bibitem{tpbc}{Periodic and twisted periodic longitudinal boundary 
conditions yield the same locus ${\cal B}$, so the ${\cal B}$ for the cyclic
and M\"obius strips of a given type are the same and separately ${\cal B}$ for
the torus and Klein bottle strips of a given type are the same
\cite{pm,tk,tor4}.}
\end{thebibliography}
\end{document}